\documentclass[3p]{elsarticle}

\usepackage{graphicx}
\usepackage{wrapfig}
\usepackage{amsmath,amssymb,fontenc, mathptmx}

\begin{document}

\def\lb{\linebreak[4]}
\newcommand{\be}{\begin{equation}}
\newcommand{\ee}{\end{equation}}
\newcommand{\bea}{\begin{eqnarray}}
\newcommand{\eea}{\end{eqnarray}}
\newcommand{\bes}{\begin{subequations}}
\newcommand{\ees}{\end{subequations}}
\newcommand{\bear}{\begin{equation}\begin{array}}
\newcommand{\eear}[1]{\end{array}\label{#1}\end{equation}}
\newcommand{\fr}[2]{\dfrac{{ #1}}{{ #2}}}
\newcommand{\pa}{\partial}
\newcommand{\la}{\langle}
\newcommand{\ra}{\rangle}
\newcommand{\fn}[1]{\footnote{{#1}}}
\newcommand{\bu}{$\bullet$\ }
\renewcommand{\le}{\leqslant}
\renewcommand{\ge}{\geqslant}
\def\vak{{\varkappa}}
\def\vep{{\varepsilon}}
\newcommand{\epe}{\mbox{$e^+e^-\,$}}
\newcommand{\ggam}{\mbox{$\gamma\gamma\,$}}
\newcommand{\egam}{\mbox{$e\gamma \,$}}
\newcommand{\eeww}{\mbox{$e^+e^-\to W^+W^-\,$}}
\newcommand{\ggww}{\mbox{$\gamma\gamma\to W^+W^-\,$}}
\newcommand{\ggzz}{\mbox{$\gamma\gamma\to ZZ\,$}}
\newcommand{\mup} {\mbox{$\mu^+\mu^-\,$}}
\newcommand{\DDp}{\mbox{$D^+D^-\,$}}
\def\cl{\centerline}

\newenvironment{Itemize}{\begin{list}{$\bullet$}%
{\setlength{\topsep}{0.2mm}\setlength{\partopsep}{0.2mm}%
\setlength{\itemsep}{0.2mm}\setlength{\parsep}{0.2mm}\setlength{\leftmargin}{4mm}}}%
{\end{list}}
\newcounter{enumct}
\newenvironment{Enumerate}{\begin{list}{\arabic{enumct}.}%
{\usecounter{enumct}\setlength{\topsep}{0.2mm}%
\setlength{\partopsep}{0.2mm}\setlength{\itemsep}{0.2mm}%
\setlength{\parsep}{0.2mm}\setlength{\leftmargin}{4mm}}}
{\end{list}}

\newenvironment{fmpage}[1]
{\begin{lrbox}{\fmbox}\begin{minipage}{#1}}
{\end{minipage}\end{lrbox}\fbox{\usebox{\fmbox}}}
\newcommand{\missET}{\slash{\hspace{-2.4mm}E}_T}

\date{}

\title{Beam Dump problem and Neutrino Factory Based on a
$e^+e^-$ Linear Collider}

\author{I. F. Ginzburg}
\ead{ginzburg@math.nsc.ru}
\address{
Sobolev Institute of Mathematics, Prosp.ac. Koptyug,4\\ and Novosibirsk State University, ul. Pirogova, 2\\
 Novosibirsk, 630090 Russia}

\sep



\begin{abstract}
The beam of  an $e^+e^-$ Linear Collider after a collision
at the main interaction point  can be
utilized to construct   the  neutrino factory with exceptional
parameters. We also discuss briefly possible applications
of some elements of  the  proposed scheme to   standard fixed
target experiments and new experiments with $\nu_\mu N$
interactions.
\end{abstract}
\maketitle

\section{Introduction}

The projects of  the  $e^+e^-$ Linear Collider (LC) -- ILC, CLIC, ... -- contain one essential element that is not present in other colliders.
Here each $e^-$ (or $e^+$) bunch will be used only
once, and physical collisions retain two very dense and
strongly collimated  beams of high energy electrons  with  the
precisely known time structure. We consider,
for definiteness, electron beam parameters of the ILC
project \cite{ILC}:
  \bear{l}
\mbox{\it particle energy}\;\; E_e=250\div 500\;{\rm GeV}, \\
\mbox{\it number of electrons per second}\;\; N_e\sim 10^{14}/{\rm s},\\
\mbox{\it transverse size and angular spread are negligible};\\
\mbox{\it time structure is complex and precisely known}.
 \eear{beampar}

The problem of dealing with this powerful beam dump is
under extensive discussions, see, e.g.,~\cite{ILC}.

About 10 years ago we suggested to utilize such used beams  in project TESLA to
initiate operation of a subcritical fission reactor and to
construct  a  neutrino factory ($\nu$F)~\cite{LCWS05,reactor}.
With new studies of ILC and CLIC,  these proposals should be renewed.\\

\bu Neutrino factories promise to solve many problems. The existing
projects (see, e.g.,~\cite{nufact1}-\cite{nufact3}) are very
expensive, their physical potential is limited by an
expected neutrino energy and productivity of a neutrino
source.

The proposed $\nu$F based on LC is much less expensive
than those discussed nowadays since there are no additional costs
for construction of a  high intensity  and  high energy particle source. The
combination of the high number of particles in the beam and
high particle energy with precisely known time
structure \eqref{beampar} provides very favorable properties of such a $\nu$F.
The initial beam will be prepared in LC irrelevantly to the $\nu$F
construction. The construction demands no special
electronics except  that for detectors. The initial beam
is very well collimated, therefore the additional efforts for
beam cooling are not necessary. Use of the IceCube in
Antarctica or Lake Baikal detector just as specially prepared detector not so far from LC as a far distance detector (FDD)
allows to study in details $\nu_\mu-\nu_\tau$ oscillations and observe possible oscillations $\nu_\mu\to  \nu_{sterile}$ (in latter case --
via a measurement of deficit of $\nu_\mu N\to \mu X$ events).

The neutrino beam will have a very well known discrete time
structure  that repeats the same structure in the LC. This
fact allows one to separate cosmic and similar backgrounds
during operation with high precision. A very simple structure of
a neutrino generator allows to calculate the energy spectrum
and  a  content of the main neutrino beam with high accuracy.
It must be verified with high precision in a nearby detector
(NBD).

In this project an incident neutrino beam will contain mainly $\nu_\mu$,
$\bar{\nu}_\mu$ with a small admixture of $\nu_e$
and $\bar{\nu}_e$ and tiny dope of $\nu_\tau$ and
$\bar{\nu}_\tau$ (the latter can be calculated with low
precision). For the electron beam energy of 250~GeV, neutrino
energies are spread  up to about
80~GeV with the  mean energy of about 30 GeV, providing reliable
observation of $\tau$, produced by $\nu_\tau$ from the
$\nu_\mu-\nu_\tau$ oscillations. In the physical program of a
discussed $\nu$F we consider  only the problem of
$\nu_\mu-\nu_\tau$ and/or $\nu_\mu
-\nu_{sterile}$ oscillations. The potential of this $\nu$F in  solving
other problems of $\nu$ physics should be studied after
detailed consideration of the project, see also~\cite{nufact3}.\vspace{-3mm}

\section{Elements of neutrino factory. Scheme}


The proposed scheme deals with the electron beam used in LC
and contains the following parts,  Fig.~1:\\
\bu Beam bending magnet (BM).\\
\bu Pion producer (PP), \, \\ \bu Neutrino
transformer (NT),\\
 \bu Nearby detector (NBD), \\ \ \bu Far distance detector (FDD).
\\
\begin{figure}[hbt]
\centering \includegraphics[width=0.75\textwidth, height=2.5cm]{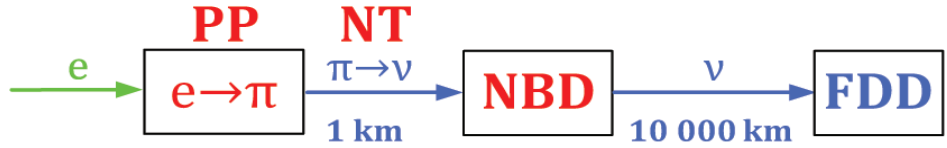}
\caption{Main parts of the neutrino factory after BM.}
 \end{figure}

\section{Beam bending magnet}

The system should start with a bending magnet situated after
the detector of the basic collider.  It
turns the used beam at an angle necessary to reach FDD  by
sacrificing  monochromaticity but without essential growth of the angular
spread. The vertical component of the turning angle $\alpha_V$
is determined by Earth curvature. Let us denote the
distance from LC to FDD at the Earth surface by $L_F$. To reach FDD,  the initial
beam (and therefore NT) should be turned before PP at the
angle $\alpha_V =L_F/(2R_E)$ below horizon  (here
$R_E$ is the Earth radius).

The horizontal component of turning angle can be minimized
by a suitable choice of the proper LC orientation
(orientation of an incident beam near the LC collision point).

\section{Pion producer (PP)}

The PP can be, for example, a 20~cm long water cylinder
({\it one radiation length}). Water in the cylinder can rotate for cooling.
In this PP almost each electron will produce a bremsstrahlung
photon with energy $E_\gamma=100-200$~GeV. The angular
spread of these photons is roughly the same as that
of the initial beam ($\sim 0.1$~mrad). Bremsstrahlung
photons have an additional angular spread of about
$1/\gamma\approx 2\cdot 10^{-6}$. These two spreads are
negligible for our problem.

Then these photons collide with nuclei and produce pions,
 \be
\gamma N\to N'+\pi\;'s,\quad \sigma\approx 110\,\mu b.
 \ee
This process gives about $10^{-3}$ $\gamma N$ collisions
per one electron that corresponds to about $\sim 10^{11}$ $\gamma N$
collisions per second. On average, each of these collisions
produces a single pion with high energy $E_\pi>E_\gamma/2$ (for
estimates $\la E_\pi^h\ra=70$~GeV) and at least 2-3 pions
with lower energy (for estimates, $\la E_\pi^\ell\ra\approx
20$~GeV).

Mean transverse momentum of these pions is 350-500 MeV. The
angular spread of high energy pions with  the energy $\la
E_\pi^h\ra$  is within 7 mrad. Increase of the angular
spread of pions with decrease of energy is compensated by
growth of the number of produced pions. Therefore, for
estimates we accept that the pion flux within an angular
interval of 7~mrad contains $\sim 10^{11}$~pions with
$E_\pi=\la E_\pi^h\ra$ and the same number of pions with
$E_\pi=\la E_\pi^\ell\ra$ per second. Let us denote the
energy distribution of  pions flying almost forward by
$f(E)$.

$\circ$ Certainly,  more refined calculations  should also consider
production and decay of $K$ mesons etc.

\bu The production of $\nu_\tau$ in the reaction mentioned in ref.~\cite{Telnov}
 \begin{subequations}\label{nutsorce}
 \be
\gamma N\to D_s^\pm X\to \nu_\tau \bar{\tau} X\,.
 \ee
plays the most essential role for our estimates. Its cross
section rapidly increases with energy growth and
 \be
\sigma(\gamma N\to \tau+...)\approx 2\cdot 10^{-33}\, {\rm cm}^2\;\;
{\rm at} \;\;
E_\gamma\approx 50 \mbox{ GeV}\,.
 \ee
 \end{subequations}

\section{Neutrino transformer (NT). Neutrino beams}

For the neutrino transformer (NT), we suggest a high vacuum
beam pipe of length $L_{NT}\approx 1$~km and radius
$r_{NT}\approx 2$~m.
Here muon neutrino $\nu_\mu$ and
$\bar{\nu}_\mu$ are created from $\pi\to\mu\nu$ decay. This
length $L_{NT}$ allows that more than one quarter of pions with
$E_\pi\le\la E_\pi^h\ra$ decay. The pipe with a radius
$r_{NT}$ gives an angular coverage of 2 mrad, which cuts
out 1/12  of the total flux  of low and medium energy
neutrinos. With the growth of pion energy, two factors act
in opposite directions. First,  the initial angular
spread of pions decreases with this growth, therefore, the fraction
of the flux selected by the pipe  will increase. Second,  the
number of pion decays within a relatively short pipe
decreases  with this growth. These two tendencies compensate each
other in the resulting flux.

The energy distribution of neutrinos obtained from the decay of a pion
with the energy $E$ is uniform in the interval $(aE,\,0)$
with $a=1-(m_\mu /m_\pi)^2$. Therefore, the
distribution $F(\vep)$ of the neutrino energy $\vep$  can be obtained from
the energy distribution of pions near the forward direction $f(E)$
as
 \be
F(\vep)=\int\limits_{\vep/a}^{E_e} f(E)dE/(aE)\,,\qquad
a=1-\fr{m_\mu^2}{m_\pi^2}\approx 0.43\,.\label{spectr}
 \ee
The increase of the angular spread in the decay is negligible
in the rough approximation. Finally, at the end of NT we
expect to have the neutrino flux within the angle of 2~mrad
 \bear{c}
2\cdot 10^{9} \nu/{\rm s} \;\; {\rm with}\;\; E_\nu=\la
E_\nu^h\ra\approx 30~{\rm GeV},\\ \mbox{ and }\;\; 2\cdot
10^{9} \nu/{\rm s} \;\; {\rm with}\;\; E_\nu=\la E_\nu^\ell\ra\approx
9~{\rm GeV}.
 \eear{nucount}
We denote below neutrinos with $\la E_\nu\ra=30$~GeV and
$9$~GeV as {\it high energy neutrinos} and  {\it low
energy neutrinos}, respectively.

$\circ$ Other sources of $\nu_{\mu}$ and $\nu_e$ change these
numbers only slightly.

\bu {\bf The background $\pmb{\nu_\tau}$ beam}.

The $\tau$ neutrino are produced  in PP. Two mechanisms
were discussed in this respect, the Bethe-Heitler process
$\gamma N\to \tau\bar{\tau}+X$~\cite{Skrinsky} and the process
\eqref{nutsorce} which is  dominant~\cite{Telnov}. The
cross section \eqref{nutsorce} is five orders smaller than
$\sigma(\gamma N\to X)$. The mean transverse momentum $\la p_t\ra$ of $\nu_\tau$ is given by $m_\tau$, it is more than three times
higher than $\la p_t\ra$ for $\nu_\mu$. Along with, e.g.,
$\bar{\nu}_\tau$ produced in this process, in NT each
$\tau$ decays to $\nu_\tau$ plus other particles.
Therefore,  each  reaction  of such a type is a source of a
a $\nu_\tau+\bar{\nu}_\tau$ pair. Finally, for the flux density
we have
 \be
N_{\nu_\tau}\sim 3\cdot 10^3\nu_\tau/({\rm s}\cdot {\rm mrad}^2)\lesssim
8\cdot 10^{-6} N_{\nu_\mu}\,.\label{nutflux}
 \ee
The $\nu_\tau$ (or $\bar{\nu}_\tau$) energy is typically
higher than that of $\nu_\mu$ by  a factor of $2\div 2.5$.

Besides,  $\nu_\tau$ will be produced by non-decayed pions
within the protecting wall behind NP in the process like
$\pi N\to D_sX\to \tau\nu_\tau X$. The cross section of
this process increases  rapidly with  the energy growth and
equals $0.13\mu$b at $E_\pi=200$~GeV \cite{tel1}.   A rough
estimate shows that the number of additional $\nu_\tau$
propagating in the same angular interval is close to the
estimate \eqref{nutflux}. In the numerical
estimates below we consider, for definiteness, the first
contribution only. A measurement of $\nu_\tau$ flux in
the NBD is a necessary component for the study of $\nu_\mu-\nu_\tau$
oscillations in FDD.

\section{ Nearby detector (NBD)}

The main goal of the nearby detector (NBD) is to measure
the energy and angular distribution of neutrinos within
the beam as well as $N_{\nu_e}/N_{\nu_\mu}$ and
$N_{\nu_\tau}/N_{\nu_\mu}$.

We propose to place the NBD at the reasonable distance behind NT
and a concrete
wall (to eliminate pions and other particles from the initial
beam).  For estimates, we consider the body of NBD in a form of the
water cylinder with a radius about 2-3~m (roughly the same as
NT) and length $\ell_{NBD}\approx 100$~m. The detailed construction
of the detector should be considered separately.

For $E_\nu=30$ GeV, the cross section for $\nu$ absorption
is
 \bear{c}
 \sigma(\bar{\nu} N\to \mu^+h)=0.1 \pi\alpha^2\fr{m_pE_{\bar{\nu}}}
{M_W^4\sin^4\theta_W}
 \approx
10^{-37} {\rm cm}^2,\\ \sigma(\nu N\to
\mu^-h)=0.22 \pi\alpha^2\fr{m_pE_\nu}{M_W^4\sin^4\theta_W}\approx 2\cdot 10^{-37}
{\rm cm}^2.
 \eear{}
Taking into account  these numbers,  the free path length in water is
 $\lambda_{\bar{\nu}}=10^{13}$~{\rm cm} and $\lambda_\nu= 0.45\cdot
10^{13}$~{\rm cm}. That gives
 \bear{c}
(1\div 2)\cdot 10^7\;\; \mu/{ {\rm year}}\;\;
 ({\rm with}\;\;\la E_\mu\ra\sim 30\;{\rm GeV});\\
150\div 250\;\; \tau/{ {\rm year}}\;\;  ({\rm with}\;\;\la
E_\tau\ra\sim 50\;{\rm GeV})\,.
 \eear{numberNBD}
(here 1 year =$10^7$~s, that is LC working time). These numbers look sufficient for
detailed measurements of muon neutrino spectra and for
verification of   the calculated direct $\nu_\tau$ background.

\section{Far Distance Detector (FDD)}

Here we consider how the  FDD  can be used for a solution of the single problem:
$\nu_\mu-\nu_\tau$ and (or) $\nu_\mu-\nu_{sterile}$ oscillations.
Other possible applications should be considered elsewhere.
We discuss here two possible position of FDD -- at relatively small distance from LC -- FDD I (with special detector) and very far from LC -- FDD II (with using big detectors, constructed for another goals).

For the
length of oscillations we use  estimate  \cite{Vysot}
 \be
 L_{osc}\approx
E_\nu/(50~{\rm GeV})\cdot 10^5~{\rm km}\,.\label{Losc}
 \ee

\subsection{FDD I}

We discuss first the opportunity to construct special relatively compact detector with not too expensive excavation work  at the distance  of a few hundred kilometers from LC (for definiteness, 200 km). For this distance the  NT should be
turned at 16~mrad  angle below horizon. This angle can be reduced by
3~mrad (one half of angular spread of initial pion beam).

We consider the body of this FDD in the form of water channel of length 1~km with radius $R_F\approx 40$~m. The transverse size is
limited by water transparency.

The fraction of neutrino's reaching this FDD is given by
ratio $k=(R_F/L_F)^2/[(r_{NT}/L_{NT})^2]$. In
our case $k\approx 0.01$. Main effect under interest here is
$\nu_\mu\to \nu_\tau$ oscillation.  They add
$(L_F/L_{osc})^2N_{\nu_\mu}$ to initial $N_{\nu_\tau}$.

In FDD of chosen sizes we expect the counting rate to be
just 10 times lower than that in NBD \eqref{numberNBD} for
$\nu N\to \mu X$ reactions with high energy neutrino. We
also expect the rate of $\nu_\tau N\to \tau X$ events to be
another $10^5$ times lower (that is about 10 times higher
than the background given by initial $\nu_\tau$ flux),
  \be
\begin{array}{c}
  N(\nu_\mu N\to \mu X)\approx (1\div 2)\cdot
  10^6/year,\\
N(\nu_\tau N\to \tau X)\approx (10\div
20)/year\end{array}\;\; in\;\; FDDI.\label{FDDInumb}
 \ee
For neutrino of lower energies effect  increases. Indeed,
$\sigma(\nu N\to\tau X)\propto E_\nu$ while $L_{osc}\propto
E_\nu$. Therefore,  observed number of $\tau$ from
oscillations increases $\propto 1/E_\nu$ at $E_\nu\ge
10$~GeV. The additional counting rate for $\nu_\tau N\to
\tau X$ reaction with low energy neutrino (with $\la
E_\nu\ra=9$~GeV) cannot be estimated so simply, but rough
estimates give   numbers similar to \eqref{FDDInumb}.

These numbers look sufficient for separation of
$\nu_\mu-\nu_\tau$ oscillations and rough measurement of
$s_{23}$.

Note that at considered FDD I size the counting rate of $\nu_\tau
N\to \tau X$ reaction is independent on FDD distance from
LC, $L_F$. The growth of $L_F$ improves the signal to
background ratio for oscillations. The value of signal
naturally increases with growth of volume of
FDD I.

\subsection{FDD  II}

Now we consider very attractive opportunity to use for FDD existent neutrino telescope with volume of 1 km$^3$ situated at Lake Baikal or in Antarctica (IceCube detector) with the
distance {\it basic LC --- FDD II}  $L_F\approx 10^4$~km. This
opportunity requires  an  excavation work
for NT and NBD at the angle about $50^\circ$ under horizon.

At this distance according to  \eqref{Losc} for $\nu$ with energy about 30~GeV, we expect
the conversion of $(L_F/L_{osc})^2\approx 1/36$ for
$\nu_\mu\to \nu_\tau$ or $\nu_\mu\to \nu_{sterile}$.

The number of expected events $\nu_\mu\to \mu
X$ with high energy neutrinos will be about 0.01 of that in
NBD,
 \be
 \begin{array}{c}
  N(\nu_\mu N\to \mu X)\approx (1\div 2)\cdot
  10^5/{\rm year},\\
N(\nu_\tau N\to \tau X)\approx 3\cdot 10^3/{\rm year}\,.\end{array} \label{FDDIInumb}
 \ee
The contribution of low energy neutrinos increases both
these counting rates.

Therefore, one can hope that a few years of experimentation
with a reasonable $\tau$ detection efficiency will allow one to
measure $s_{23}$ with percent accuracy, and a similar period
of observation of $\mu$ production will allow to observe
the loss of $\nu_\mu$ due to transition of this neutrino to
 $\nu_{sterile}$.

\section{Discussion }

Here we suggest to construct Neutrino Factory  with great physical potential using beam of Linear Collider after main collision there.

All technical details of the proposed scheme including
sizes of all elements, construction, and materials of
detectors can be modified in the forthcoming simulations
and optimization of parameters. The numbers obtained above
represent the first careful estimates. In particular, rate of neutrino can be  enhanced with increasing of length of PP, this rate can appear significantly higher after more accurate calculation of pion production there, the length of NT can be reduced due to economical reasons, etc.

After first stage of Linear Collider its energy can be increased. For these stages proposed scheme can be used without changes (except magnetic field in BM and taking into account new time structure of neutrino beam).

We did not discuss here methods of $\mu$ and $\tau$ detection
and their efficiency. Next, a  large fraction of residual
electrons, photons and pions leaving the PP will reach the
walls of the NT pipe. The  heat sink and radiation
protection of this pipe must be taken into account.

A more detailed physical program of this $\nu$F
will include many features of that  in other projects
(see, e.g.,~\cite{nufact1}-\cite{nufact3}).

\section{ Other possible applications of some parts of $\pmb\nu$F}

\bu {\bf PP for the fixed
target experiment}.  The PP can be treated as an $eN/\gamma N$
collider with luminosity $3\cdot 10^{39}$~cm$^{-2}$s$^{-1}$ with a
c.m.s. energy of about 23~GeV. Therefore, if one adds some standard
detector equipment behind PP, it can be also used for a fixed target
$eN/\gamma N$ experiment.  Here one can
study  rare processes in $\gamma N$ collisions, $B$
physics, etc.

\bu {\bf Additional opportunity for using NBD.} The high rate
of $\nu_\mu N\to \mu X$ processes
expected in NBD allows one to study  new problems of high
energy physics. The simplest example is the opportunity to
study  charged and axial current induced structure functions and
diffraction ($\nu N\to \mu +hadrons$,  $\nu
N\to\mu N' \rho^\pm $, $\nu N\to\mu N' b_1^\pm$,...)
with high precision.\\

I am thankful to  N.~Budnev, S.~Eidelman, D.~Naumov, L.~Okun, V.~Saveliev,
A.~Sessler, V.~Serbo, A.~Skrinsky, O.~Smirnov, V.~Telnov, M.~Vysotsky, M.~Zolotorev for comments.
The paper is supported in part by Russian grants RFBR, NSh,
of Presidium  RAS  and
Polish Ministry of Science and Higher
Education Grant  N202 230337

\end{document}